MyShake: Detecting and characterizing earthquakes with a global smartphone seismic network

Authors: Qingkai Kong, Sarina Patel, Asaf Inbal, Richard M Allen

Corresponding author: Qingkai Kong

Address: 209 McCone Hall, UC Berkeley, Berkeley, CA, 94720

Email: kongqk@berkeley.edu

Abstract:

MyShake harnesses private/personal smartphones to build a global seismic network. It uses the accelerometers embedded in all smartphones to record ground motions induced by earthquakes, returning recorded waveforms to a central repository for analysis and research. A demonstration of the power of citizen science, MyShake expanded to 6 continents within days of being launched, and has recorded 757 earthquakes in the first 2 years of operation. The data recorded by MyShake phones has the potential to be used in scientific applications, thereby complementing current seismic networks. In this paper: (1) we report the capabilities of smartphone sensors to detect earthquakes by analyzing the earthquake waveforms collected by MyShake.  (2) We determine the maximum epicentral distance at which MyShake phones can detect earthquakes as a function of magnitude.  (3) We then determine the capabilities of the MyShake network to estimate the location, origin time, depth and magnitude of earthquakes. In the case of earthquakes for which MyShake has provided 4 or more phases (21 events), either P- or S-wave signals, and has an azimuthal gap less than 180 degrees, the median location, origin time and depth errors are 2.7 km, 0.2 s, and 0.1 km respectively relative to USGS global catalog locations.  Magnitudes are also estimated and have a mean error of 0.0 and standard deviation 0.2.  These preliminary results suggest that MyShake could provide basic earthquake catalog information in regions that currently have no traditional networks. With an expanding MyShake network, we expect the event detection capabilities to improve and provide useful data on seismicity and hazards.

**Introduction**

After more than a century of development, geophysical instrumentation has become more and more diversified. High quality seismic instruments (Havskov and Alguacil, 2016), geodetic instruments (Larson, 2009), and interferometric synthetic aperture radar (Bürgmann *et al.*, 2000) enable new discoveries and understanding of earthquake physics and active tectonics. In addition, the emergence of various new low-cost and potentially more pervasive sensing technologies provide new ways of detecting earthquakes, collecting additional data to learn about the earthquake process, and potentially, making important contributions to seismology (Allen, 2012).

"Did You Feel It," a USGS earthquake survey platform, collects macroseismic intensity data from Internet users which is then used to generate intensity maps immediately following earthquakes (Wald *et al.*, 2001; Wald and Dewey, 2005; Atkinson and Wald, 2007). Twitter messages from users who felt an earthquake can be used to detect and characterize events in real-time (Earle, 2010; Earle *et al.*, 2010; Sakaki *et al.*, 2010). By monitoring traffic to its website, the European-Mediterranean Seismological Centre can detect and assess the effect of an earthquake within a few minutes (Bossu *et al.*, 2012). Low-cost MEMS sensors inside computers or placed in specially installed stand-alone boxes in homes or offices can be used to monitor and study earthquakes (Cochran *et al.*, 2009; Chung *et al.*, 2011; Clayton *et al.*, 2012, 2015; Hsieh *et al.*, 2014; Wu, 2015; Wu *et al.*, 2016). Distributed acoustic sensing (DAS) transforms telecommunication fiber-optic cables into seismic arrays, enabling meter-scale recording over kilometers of linear fiber length (Dou *et al.*, 2017; Lindsey *et al.*, 2017).

As more and more people have access to, and a need for smartphones, these small devices comprise an ever more widespread and dense sensing network around the globe. Seismologists have learned that smartphones can be used in different ways to detect earthquakes. For example, by monitoring when users turn on a specific earthquake application on their phone, earthquakes can be recognized within minutes as clusters of application activity (Bossu *et al.*, 2015, 2018; Steed *et al.*, 2019). The MEMS sensors inside the smartphones that record acceleration have also been shown to be capable of detecting earthquakes (Faulkner *et al.*, 2011; Dashti *et al.*, 2012, 2014; Kong *et al.*, 2015; Finazzi, 2016; Kong, Allen, Schreier, *et al.*, 2016).

MyShake was launched by UC Berkeley in 2016 as a citizen science project. It aims to build a global smartphone seismic network that can be used for research, ultimately contributing to a reduction in earthquake hazards.  In the first 2 years, just under 300,000 people downloaded the MyShake app globally.  Now, two years after the launch, there are 40,000 phones with the app installed, and on any given day about 7,000 phones contribute data. The core of MyShake is an artificial neural network, built into the on-phone app, that is trained to recognize earthquake-like movement and distinguish it from everyday human movements and a series of machine learning models that support the confirmation and estimation of the earthquakes (Kong, Allen, Schreier, *et al.*, 2016; Kong, Inbal, *et al.*, 2019). Whenever the phone detects the earthquake-like movement, a real-time message with the trigger location, time and amplitude will be sent to the server for earthquake early warning purposes. At the same time a 5-minute segment of

3-component acceleration data is stored on the phone and then uploaded to the MyShake server to be analyzed when the phone connect to power and WiFi. The time series starts 1-minute before the trigger is detected, and continues for 4-minutes post-trigger. Data collected by MyShake can be used in various applications. Kong, Allen, and Schreier (2016) show examples of the waveform data recorded by MyShake, illustrating the potential to use them in different seismological applications. Real-time trigger data from MyShake users show earthquake early warning can be done using the smartphones (Kong *et al.*, 2018b). The MyShake waveform data can also potentially be used to monitor the health state of buildings (Kong *et al.*, 2018a). In addition, using the MyShake arrays, the system can potentially detect microseismicity and monitor noise in urban areas (Inbal *et al.*, 2019).

In this paper, we explore the capabilities of the MyShake smartphone seismic network by mining the archive waveforms recorded to date. Comparing this data with a global earthquake catalog, we explore the detection capabilities of the smartphone network. We explore the sensitivity of the network by determining the distance to which smartphones can detect an earthquake as a function of magnitude. We will also show how the waveforms recorded by MyShake phones can be used to estimate basic earthquake parameters, including location, origin time, depth and magnitude of the earthquake. This illustrates how the MyShake network could be used to monitor earthquake activity in regions of dense populations that currently have no seismic network.

**Data used**

The dataset used in this paper comes from global MyShake users. As described in detail in (Kong, Allen, Schreier, *et al.*, 2016), the MyShake application has a two-stage triggering algorithm. In the first stage a simple STA/LTA (short-term average and long-term average) algorithm is used to determine when a previously stationary phone moves (Allen, 1978). An ANN (artificial neural network) algorithm is used to determine whether the movement of the phone is likely due to an earthquake or other human activities. Once the movements satisfy the ANN algorithm and are determined to be earthquake-like motion, the phone will record a 5-minute segment of 3-component acceleration data that will be uploaded to the MyShake servers when the phone is next connected to WIFI and power. An earthquake waveform database is then created from the uploaded waveforms. First we scan for "candidate events" in the USGS ComCat (Comprehensive Earthquake Catalog). For each "candidate event", we search MyShake waveform archive for records within a pre-defined spatiotemporal window for possible earthquake recordings. Waveforms that meet the requirements of the above spatio temporal window are reviewed by a seismologist to filter out those caused by human activities, and to remove any bad data, e.g, missing blocks of data. Waveforms that pass all the checks are put into the earthquake waveform database. In the first two years, 757 earthquakes have had at least one recording from a MyShake user.

Figure 1a shows the location of earthquakes for which one or more seismic waveforms (confirmed by a seismologist) where uploaded from MyShake phones. Figure 1b shows the magnitude-frequency relationship of events recorded by MyShake and a comparison with the USGS ComCat catalog. For all the magnitude bins, MyShake records fewer events than the traditional seismic networks. The gap becomes smaller for larger magnitude events indicating that with increasing magnitude MyShake's capability to detect events improves, not surprisingly. Waveforms uploaded to our server are 3-component acceleration waveforms, in 5-minute segments (1-min before the trigger and 4-min after), sampled at about 25 Hz. Figure 2 shows epicentral distances of the earthquake waveforms recorded by MyShake users for earthquakes of various magnitudes. As the magnitude of the earthquakes increases, the distance from which smartphones can record useful waveforms also increases. To understand at what range we expect MyShake phones to record earthquake waveforms, we fit an analytic expression to the furthest recordings in each magnitude bin using least-square regression. For earthquakes of M2.5 to M7.8, we derived the following relationship between the magnitude of the earthquake and the distance in kilometers from which we expect to see recordings from current MyShake users:

$$R = 123e^{0.275M} - 229 \quad (2.5 \leq M \leq 7.8) \quad \text{(eq 1)}$$

Where M is the magnitude and R is the epicentral distance for the earthquake. The curve of the above equation is shown as the red curve in Figure 2. The small inset figure in Figure 2 shows the cumulative distribution of the signal to noise ratio (SNR) for all the earthquake waveforms from MyShake users. The SNR is calculated on the Y-component by selecting a 2-second window of signal centered on the peak observed ground acceleration (PGA) value, and a 2-second window of noise from the beginning of the waveform before the seismic trigger. Using the 2-second window for the signal and noise, we calculate the square of the root mean square (RMS) amplitude and take the ratio. The 25, 50, and 75 percentiles of the SNR are 6.1, 14.6 and 50.9 for the MyShake recorded earthquake waveforms.

Figure 3 shows examples of six 3-component waveforms recorded around the globe by MyShake phones. We show examples from regions that don't have dense seismic networks (additional waveforms are available in the supplementary material). The X and Y components of the acceleration records are parallel to the short and long direction of the phone-screen, while the Z component is the direction perpendicular to the phone screen. On most of these waveforms, we can see clear P- and S-waves. Because we also archive 1 minute of data before a phone trigger, even when the phone triggers on the S-phase (due to low SNR for the P-wave), we can often still observe a P-wave arrival time.

**Timing and location accuracy of smartphone records**

The accuracy of the absolute time and location associated with seismic waveforms is central to the applications and research for which the data can be used. Unlike traditional seismic stations, MyShake sensors are moving around and they do not have continuous GPS-based time. We therefore must develop strategies to improve on the normal timing and location information provided by the phone operating system to the MyShake app.

The internal phone time is insufficiently accurate as a time stamp for seismological applications of the data as it can be off by many seconds (Table 1). Instead, the MyShake app uses Network Time Protocol (NTP) to check-in with a remote server and obtain the accurate absolute time. During normal operation the MyShake app requests an NTP time stamp every hour, and the system stores this information on the backend. The statistics collected include the server time of each query, the roundtrip time for the phone's NTP query to reach and return from the server, and the offset in milliseconds of the server time relative to the phone's internal time when the query was initiated. The total actual offset between the phone's internal time and true NTP time ('true offset') was calculated as the recorded offset minus half of the roundtrip query time. We assume that it takes half of the roundtrip time for the query to reach the server.

We randomly selected 26 days from our database over a twelve-month period beginning in August of 2016. These 26 days encompass 6.4 million usable records of a successful NTP query by a phone running MyShake. Table 1 shows the distributions of these time observations. Fifty percent of queries reported an internal time better than 0.723 sec and 75% were better than 1.59 sec (true offset). The roundtrip time of an NTP time query is typically an order of magnitude smaller than this; 50% are better than 0.081 sec and 75% better than 0.131 sec.

MyShake does not attempt to correct the internal time of the phone. Instead we use the NTP times to correct the absolute timestamps associated with trigger messages and any recorded waveforms, i.e. we apply the calculated true offset time to the internal phone timestamp for the records we collect. Therefore, the accuracy of the timestamp associated with a MyShake waveform is determined by how much drift there has been to the internal phone time between the last NPT query and the time the phone triggers and records a waveform. To assess the accuracy of these timestamps, we calculate the change in the true offset values from one NTP query to the next. Figure 4 shows the distribution of the change in true offset values reported by phones during the 26 days sampled. Fifty percent of the cases have a change in true offset less than 0.027 sec, 75% are better than 0.132 sec, and 90% are better than 0.503 sec (Table 1). The magnitude of the changes increase—and therefore the accuracy of the timestamps decreases—as the time elapsed between NTP queries increases. If queries occur 1-10 minutes apart, the offset changes by <0.015 sec 50% of the time; this metric becomes <0.029 when queries of ~30 minutes apart, and <0.064 sec for ~hourly queries.

To provide location information, MyShake requests users' permission to collect GPS locations from participating phones. To assess the accuracy of smartphone GPS systems in the MyShake use case, we conducted a test approximating typical conditions for a stationary phone monitoring for an earthquake signal. In a six story building, 10 smartphones were placed on a second floor windowsill facing into a partially sheltered courtyard with a limited view of the sky. We elected to use a windowsill (rather than a location more interior to the building) so that we could determine the accurate true location of the phones by determining the location of the side of the building in Google Earth. The phones were left stationary so they could enter into 'steady' mode before being manually prompted to 'trigger', causing a seismic waveform to be recorded and the location of the phone/waveform reported as it is in normal MyShake operation. Figure 5 shows the distribution of horizontal and vertical (elevation) errors based on 98 triggers, with significant percentiles tabulated in Table 1.

For 50% of triggers, the horizontal location is within 14 m of the true location. For reference, the reported accuracy of smartphone GPS location when the phone is stationary and has a clear view of the sky is about 5 m (from GPS.gov). It is within 28.8 m 75% of the time in our test, and within 43.6 m 90% of the time (Table 1). A typical home has a footprint 10 m across, while office buildings might have a 50 m footprint. This means that the location information is of the same order as the size of buildings, and there is the potential to group waveforms by building or building type in order to both observe and correct for building amplification factors. The on-phone API providing trigger location information also provides a horizontal accuracy metric. As can been seen in Table 1, our observed errors are consistent with the reported accuracy.

The elevation is within 3.9 m of the true value 50% of the time. The typical floor spacing of a multi-story building is ~4 m. Therefore our results suggest that it is possible to estimate, within plus or minus one story, on which floor a phone was located when it recorded a waveform 50% of the time. 90% of the time, the error is within 34.1 m. This is equivalent to ~8.5 floors. This error is still small enough to allow for a qualitative estimate of whether a phone is located near the bottom, middle, or top of a tall 'skyscraper' building. Such an estimation is useful in identifying cases in which the free oscillations of a building produce an amplification effect to the signals MyShake records. The distribution of the horizontal and vertical location errors are shown in Figure 5.

**Estimating earthquake source parameters**

We use the earthquake waveforms recorded by MyShake to locate events and estimate the magnitude. In this section we focus on the accuracy of source parameter estimation for events that have seismic waveforms with good azimuthal coverage, specifically the largest azimuthal gap between stations is less than 180 degrees. This is the case for 21 events in our dataset. The results of location and magnitude estimation for all events for which we have 4 or more seismic phases detected are shown in the supplementary material.

*Methods*

First, we manually pick the P and S phases. We use Hypoinverse (Klein, 2002) to determine the location, depth, and origin time of the earthquake. Hypoinverse requires an initial location and origin time as the input. For this test, we use the geometric mean of the triggers as the initial location, and the initial origin time is set to 5 s before the first trigger time. We tested a total of 28 homogeneous velocity layer models within Hypoinverse, and found the following model in Table 2 yields good results for most of the events we test. We can only estimate the location and magnitude when there are 4 or more phase picks (either P or S or mixed) available.

To estimate the magnitude, we apply the ML relationship of (Bakun and Joyner 1984). This relation estimates local magnitude from the distance, the peak to peak amplitude, and the time span from peak to peak. As shown in (Kong, Allen, and Schreier, 2016), MyShake recordings typically have larger amplitude than free-field stations at the same epicentral distance. Therefore, we scale the PGA amplitude of the MyShake recordings by dividing the observed amplitude by a factor of 1.6. This factor is derived from all the MyShake recordings.

Figure 6 shows the results for the 21 events that have good azimuthal coverage (maximum gap in azimuthal coverage is less than 180 degrees). The median errors in the location measured as distance from the USGS catalog location, in the origin time and the depth are 2.7 km, 0.2 s, and 0.1 km respectively, with standard deviations of 2.8 km, 1.2 s, and 4.9 km. Most events have distance errors that are smaller than 5 km. The larger errors are typically for events with only a small number of phase picks. In this case, 3 out of the 5 events with distance errors larger than 5 km have only 4 picks. The MyShake magnitude estimates from these events are compared with USGS catalog magnitudes in figure 7. The mean and standard deviation of the magnitude error are 0.0 and 0.2 respectively. All these events have the errors less than 0.5 units. See figure S1 and S2 for the errors in location, origin time and magnitude of all the 44 events including the ones have limited coverages in the supplementary material.

We illustrate MyShake source parameter estimation with four events from California, Oklahoma, and Morocco (errors are shown in Table 3). The M5.2 June 10th, 2016 Borrego Springs event and the M4.4 January 4th, 2018 Berkeley event occurred in locations where a good number of MyShake users where located nearby, both events have maximum azimuthal gap less than 60 degrees (Figure 8). The distance, origin time, depth and magnitude errors for the Borrego

Spring event are 3.1 km, 0.19 s, -3.5 km, and 0.28 (estimated - catalog) respectively. Likewise, for the Berkeley event they are 1.1 km, 0.47 s, -2.4 km, and 0.37 respectively.

Figure 9 shows the results for the M5.8 September 3rd, 2016 Pawnee, Oklahoma event and the M5.6 March 15th, 2016 event offshore of Morocco. Neither event has phones nearby; all recorded waveforms are at a distance of 80 km or more from the epicenter. The azimuthal gaps in coverage are also large, being 150 and 175 degree. The source parameters are therefore not as good as the ones shown in figure 8, but are still reasonable. The distance, origin time, depth and magnitude errors for the Oklahoma event are 7.6 km, 1.74 s, 5.5 km, and -0.38 respectively. For the Morocco event they are 14.7 km, -0.46 s, 1.0 km, and -0.67.

**Discussion**

The use of smartphones as a global seismic network is still a relatively new concept for which we have to determine what the capabilities are, what information this network can provide in addition to the existing more traditional seismic networks, and what challenges the network faces. So far, the successes of the MyShake project include rapid expansion around the globe, rapid increases of instrumentation densification in the aftermath of societally significant earthquakes (both locally and globally), and longevity of the user-base as the network has settled into a relatively stable number of 40,000 users. The data collected shows potential for routine seismological applications as illustrated in this study. There are still many challenges to be explored to fully understand the capabilities of this type of network. In this discussion, we describe the difficulties, and the potential future improvements for this global smartphone seismic network.

Firstly, many elements of the data processing currently still involve human interactions. In our analysis, a seismologist needs to review the waveforms to confirm that they are useful earthquake records with a relatively high signal to noise ratio, to remove the waveforms with data issues (e.g. missing data, spikes etc.), and to manually pick the P and S wave arrivals. To build a fully functioning global seismic network, all these steps need to be automated. The very different characteristics of the network compared to traditional networks means that a new suite of processing software is needed.

Secondly, there are unique challenges to use the MyShake data. The quality of the waveforms are different in ways that are difficult to determine (Kong, Lv, *et al.*, 2019). Two different phones at the same location can have very different characteristics depending on their exact physical location, and knowing how to weight the quality of the waveforms is a challenge. There are many events for which we only detect S-wave arrivals, i.e. in the case when there are many phones but only at larger distances (~100 km or more). Therefore, developing a robust phase picking algorithm for this noisy dataset is important. In addition, phones may be located in different types of buildings, on different floors, in places where the amplification, i.e. site response, are very different. The response of the buildings, desks/furniture etc., are all factors meaning we will require the use of many phones to aggregate the results in an average sense. Further calibration of the amplitude of MyShake recordings against the traditional seismic stations will be helpful. While the data shown in this study illustrated that MyShake data can be used to detect and characterize earthquakes, improving the location, origin time and magnitude estimates will require us to better understand these aspects of waveform data quality.

Thirdly, the citizen science nature of the MyShake project brings inherent challenges to seismic network operation. The configuration and density of the smartphone network is constantly changing as users join the network and leave, and individual phones move around during the day. The detection capability is much greater at night than during the day as more phones are steady during the night. MyShake users will tend to be clustered in large cities where populations are concentrated. Therefore, earthquakes in rural areas or far away from population centers can be missed. When earthquakes are detected away from population centers the station covered has only a narrow azimuthal coverage. This is similar to the situation for regional seismic networks when the earthquake is outside the seismic network. The effect is illustrated in Figure 10 which shows the errors in location and origin time with the azimuthal coverage and number of phases used in analysis for all the events that has at least 4 identifiable phases or more to conduct the calculation. When the azimuthal coverage is limited there are larger errors. For this type of earthquake, location estimation from traditional regional seismic networks are also poor. One possible solution worth exploring is combining information from traditional seismic stations and/or complementing the network with permanent low cost sensors (Cochran *et al.*, 2009; Clayton *et al.*, 2012; Wu, 2015; Nof *et al.*, 2019).

While the citizen science nature of this project brings with it the limitations described above, it also brings the possibility of rapid network expansion as there are now almost three billion smartphones in use in the world. In order to harness a larger fraction of these accelerometers in the future, MyShake has developed a completely redesigned version of the app using a human-centered design methodology (Rochford *et al.*, 2018). The new app brings additional user functionality that we hope will expand the number of users significantly providing data for more earthquakes and further study of the scientific opportunities that MyShake could support.

**Conclusion**

MyShake has now been operating as a global smartphone seismic network for several years. While there is constant turnover in the people who are running MyShake on their phone, the number of users at any given time has settled to about 40,000 globally.  The network has recorded useful seismic waveforms for hundreds of earthquakes with magnitudes from less than two up to M7.8, and from the surface to 350 km depth.  MyShake phones can trigger on and detect M3 earthquakes out to 50km, M5 out to 250 km, M7 out to 500 km.

The accuracy of the location and timing information determines the scientific uses of the data. 50% of the sampled global dataset has timestamps with accuracies better than 0.03 s. Based on a limited test of phones inside a building, 50% of locations are within 14 m horizontally and 4 m vertically.  This makes it possible to estimate in which building and what floor a record is being recorded.

Applying standard regional seismic network techniques to the MyShake data, we can determine the ability of MyShake to characterize events.  Using a set of 21 events for which there are 4 or more P- and/or S-wave arrivals and a maximum azimuthal gat less that 180 degrees, we find that the median location, depth and origin time error to be 2.7 km, 0.1 km, and 0.2 s, respectively. The mean and standard deviation of the magnitude error are 0.0 and 0.2 respectively. When earthquakes occur beneath urban regions (where there are many MyShake phones) the locations are smaller.

These preliminary results suggest the potential of the MyShake network to contribute to the seismology community by providing additional data to detect earthquakes and constrain source characteristics. In particular, locations where there are few traditional seismic stations, but dense population, MyShake could provide valuable data to constrain earthquake hazards. There are many challenges and limitations to address and overcome, but a network such as MyShake can enhance our ability to better understand earthquakes and hazards globally, as well as to engage the public in locations where these earthquake occur.

**Data and Resources**

The USGS Comcat catalog can be accessed at: https://earthquake.usgs.gov/fdsnws/event/1/. MyShake data are currently archived at Berkeley Seismology Laboratory and use is constrained by the privacy policy of MyShake (see http://myshake.berkeley.edu/privacy-policy/index.html).

**Acknowledgements**

The Gordon and Betty Moore Foundation fund this analysis through grant GBMF5230 to UC Berkeley. We thank the MyShake team members: Roman Baumgaertner, Garner Lee, Arno Puder, Louis Schreier, Stephen Allen, Stephen Thompson, Jennifer Strauss, Kaylin Rochford, Doug Neuhauser, Stephane Zuzlewski, Sarina Patel and Jennifer Taggart for keeping this Project running and growing. All the analysis of this project is done in Python, particularly the ObsPy package (Beyreuther *et al.*, 2010; Wassermann *et al.*, 2013; Krischer *et al.*, 2015). The ANSS catalog data for this study were accessed through the Northern California Earthquake Data Center (NCEDC), doi:10.7932/NCEDC. We also thank all the MyShake users who contribute to the project! In the analysis of the data, we thank Dr. Taka'aki Taira to provide helpful discussion about the use of hypoinverse.

**Tables, with captions above each table**

Table 1. Summary of the location and timing accuracy for MyShake triggers and waveforms.

|  | 50th percentile | 75th percentile | 90th percentile |
| --- | --- | --- | --- |
| Offset of internal phone clock (sec) | 0.723 | 1.59 | 20.8 |
| NTP query roundtrip time length (sec) | 0.081 | 0.131 | 0.251 |
| Accuracy of waveform and trigger timestamps (sec) | 0.027 | 0.132 | 0.503 |
| Measured horizontal location error (m) | 14.0 | 28.8 | 43.6 |
| Reported horizontal location accuracy (m) | 19.5 | 20.8 | 45.3 |
| Measured elevation error (m) | 3.9 | 11.4 | 34.1 |

Table 2. Velocity model used in the estimation of the location, origin time and depth of the earthquake using the manually picked P-wave and S-wave arrivals (except for the M5.8 September 3th, 2016 Oklahoma event, for which we use the velocity model described in (Grandin *et al.*, 2017))

| Top depth of the layer (km) | P-velocity (km/s) | S-velocity (km/s) |
| --- | --- | --- |
| 0.0 | 3.57 | 2.04 |
| 1.5 | 5.35 | 3.06 |
| 5.1 | 5.83 | 3.33 |
| 15.0 | 6.86 | 3.92 |
| 29.0 | 7.95 | 4.54 |

Table 3. Catalog and estimation source parameters with the errors for the 4 different earthquakes.

| Event name | | Origin time | Location | Depth | Magnitude |
|---|---|---|---|---|---|
| Borrego Springs, CA, USA | Catalog | 2016-06-10 08:04:38.700 | Lat: 33.43 Lon: -116.44 | 12.3 km | 5.17 |
| | Estimation | 2016-06-10 08:04:38.890 | Lat: 33.45 Lon: -116.47 | 8.8 km | 5.45 |
| | Error | 0.19 s | 3.1 km | -3.5 km | 0.28 |
| Berkeley, CA, USA | Catalog | 2018-01-04 10:39:37.730 | Lat: 37.86 Lon: -122.26 | 12.3 km | 4.38 |
| | Estimation | 2018-01-04 10:39:38.200 | Lat: 37.84 Lon: -122.26 | 9.9 km | 4.75 |
| | Error | 0.47 s | 1.1 km | -2.4 km | 0.37 |
| Pawnee, OK, USA | Catalog | 2016-09-03 12:02:44.310 | Lat: 36.43 Lon: -96.93 | 5.4 km | 5.8 |
| | Estimation | 2016-09-03 12:02:46.050 | Lat: 36.47 Lon: -97.00 | 10.9 km | 5.42 |
| | Error | 1.74 s | 7.6 km | 5.5 km | -0.38 |
| Offshore Morocco | Catalog | 2016-03-15 04:40:40.020 | Lat: 35.76, Lon: -3.61 | 10.0 km | 5.6 |
| | Estimation | 2016-03-15 04:40:39.560 | Lat: 35.63, Lon: -3.67 | 11.0 km | 4.93 |
| | Error | -0.46 s | 14.7 km | 1.0 km | -0.67 |

**Figures, with captions below each figure**

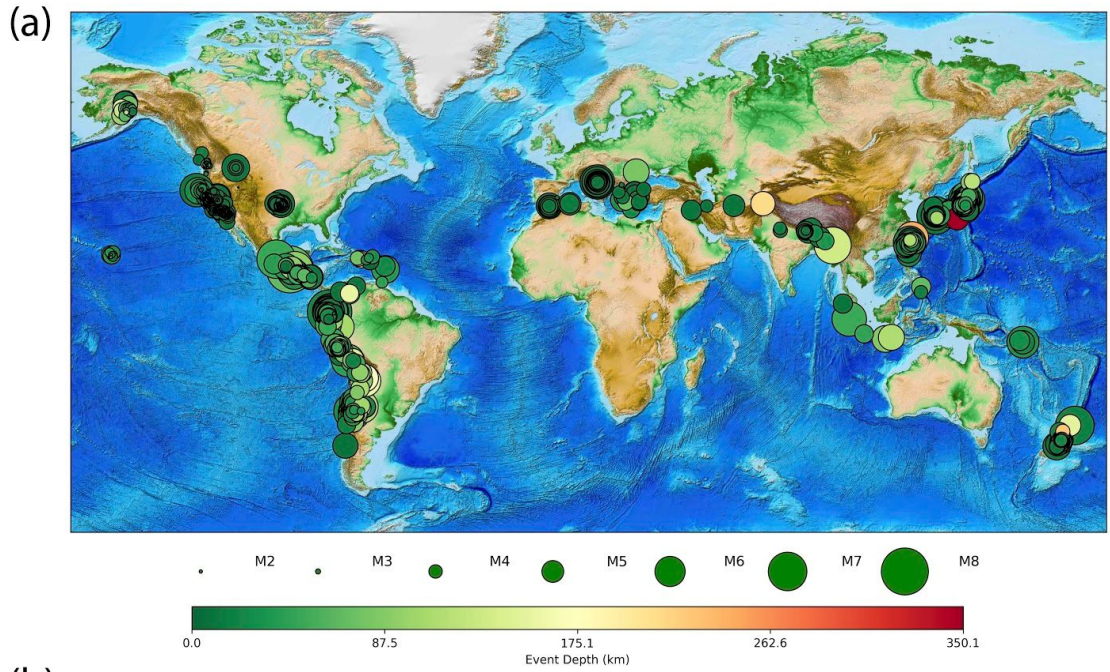

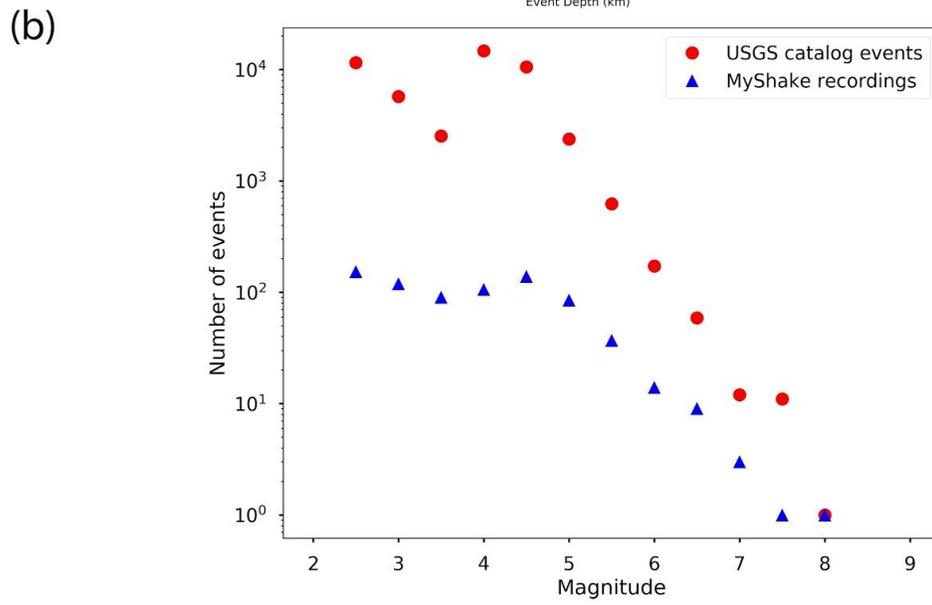

Figure 1. (a) Earthquakes with one or more useful waveform recordings from MyShake phones in the first two years of operation (Feb 12, 2016 to Feb 12, 2018). The size of the circle and color represent magnitude and depth of the earthquake (both magnitudes and locations are from USGS ComCat catalog). (b) Earthquake magnitude-frequency relation for earthquakes detected by MyShake (blue triangles) and in the USGS catalog (red circle). The number of events is measured in 0.5 magnitude bins. Figure S1 shows the difference of the number of events recorded by MyShake and the USGS catalog on the map.

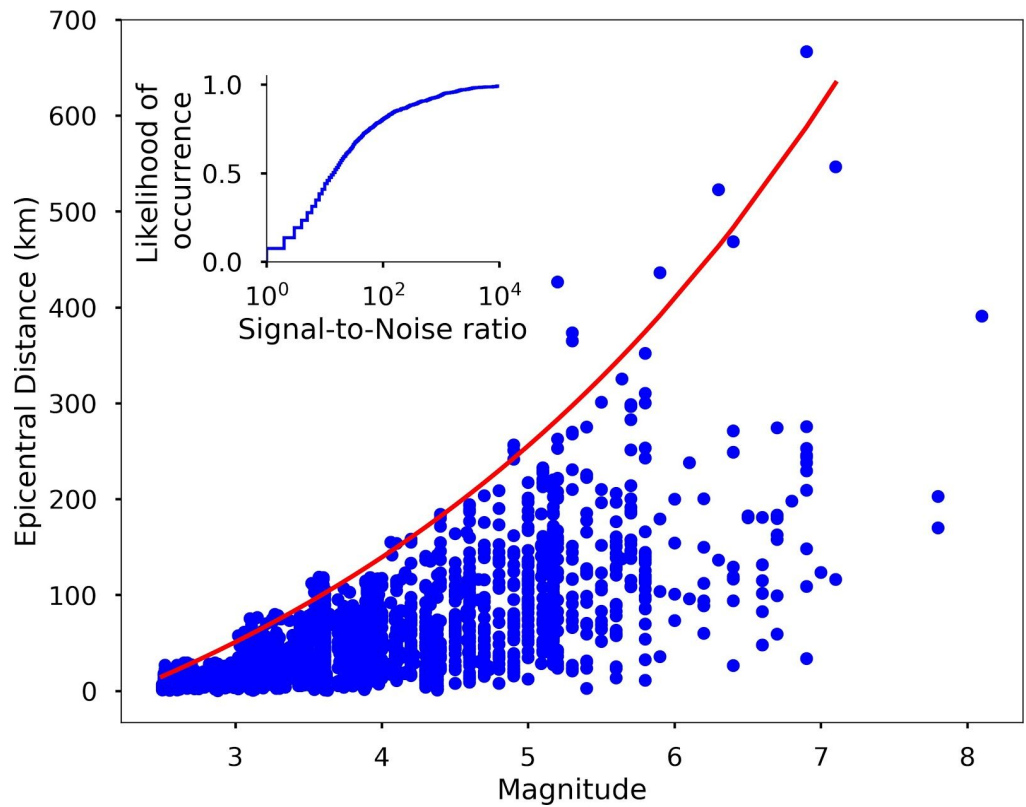

Figure 2. Epicentral distance of all MyShake earthquake waveform recordings as a function of magnitude (blue dots). The red curve is equation (1) which approximates the maximum distance to which MyShake phones can trigger and detect earthquakes. We only searched for waveforms corresponding to earthquakes M2.5 and greater in the USGS catalog. The inset figure on the top left is the cumulative distribution of the signal to noise ratio (SNR) for all the earthquake recordings measured on one horizontal component (Y component).

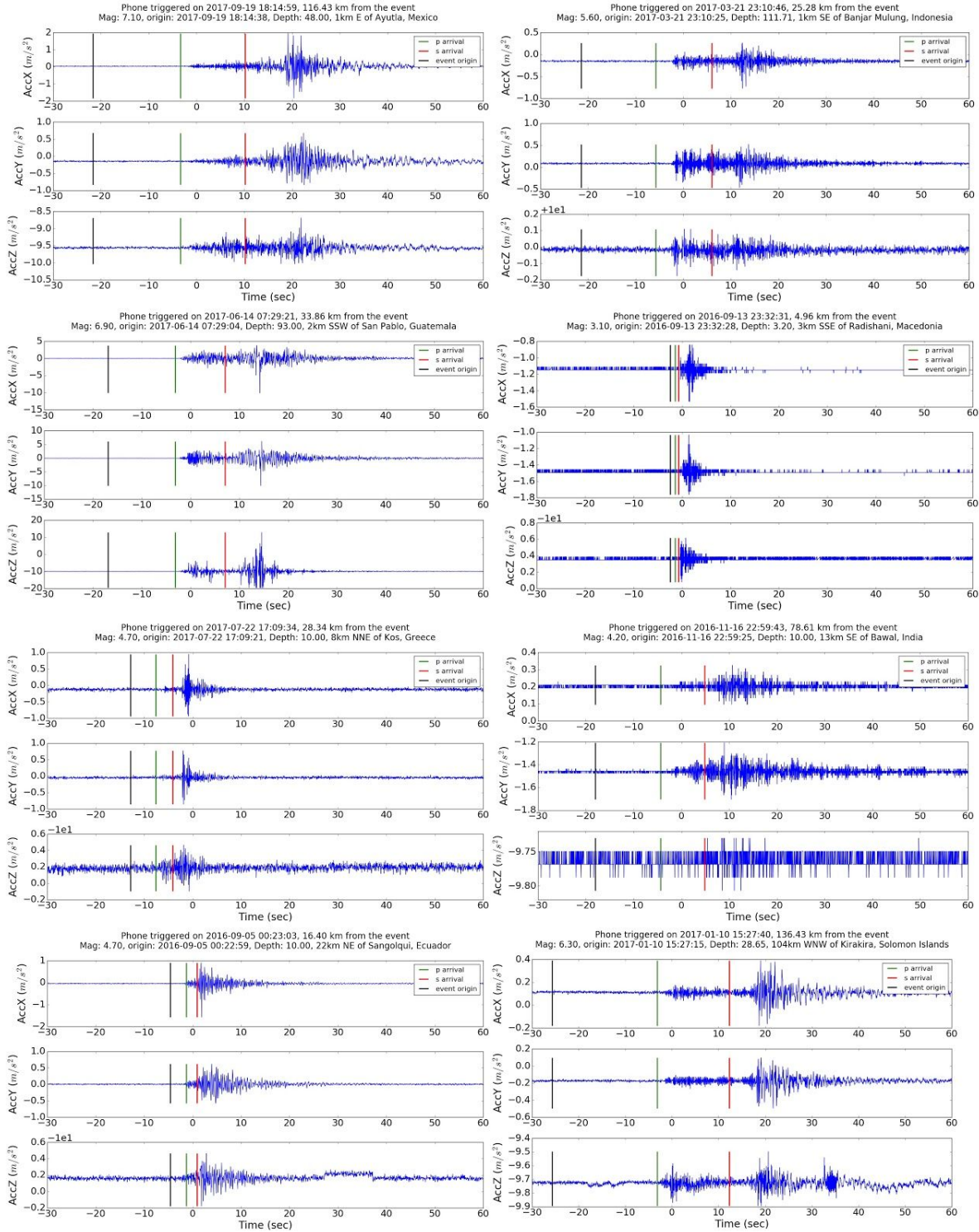

Figure 3. Example 3-component acceleration waveforms from MyShake detections globally. The black line is the event origin time from USGS catalog, green and red lines are estimated P and S arrival time using ak135 (Kennett *et al.*, 1995). The zero time on each panel is the time when the phone triggers.

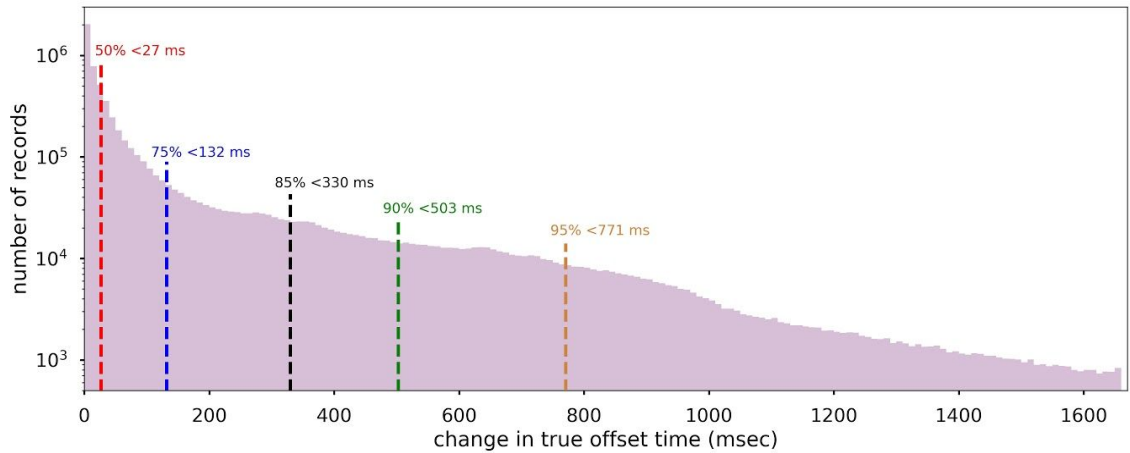

Figure 4. Accuracy of MyShake waveform and trigger timestamps. The histogram shows the change in phone timing offsets extracted from the NTP synchronization process. Data are from 26 random days between Aug 2016 to Aug 2017. The dashed lines mark the 50th, 75th, 85th and 90th, and 95th percentiles of the distribution. Note that the vertical scale is logarithmic.

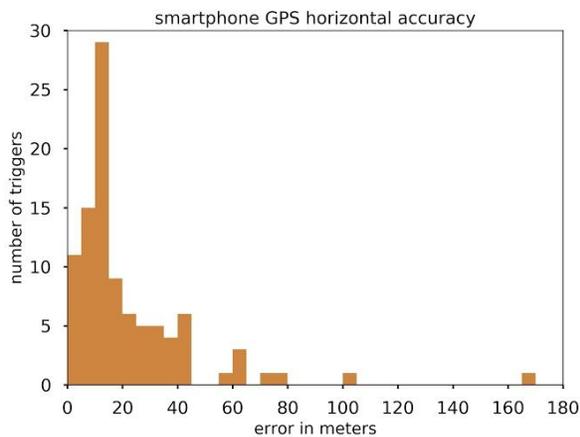
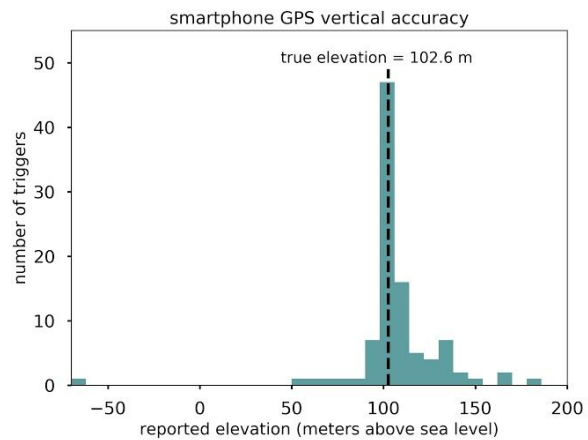

Figure 5. Accuracy of smartphone location information GPS points reported with in MyShake triggers and seismic waveforms, both horizontal (left) and vertical (right). Ten10 phones were placed on a second floor windowsill facing into a partially sheltered courtyard, and periodically prompted to collect a spontaneous trigger and record a waveform. The resulting 98 GPS points cluster closely around the true location, both horizontally and vertically.

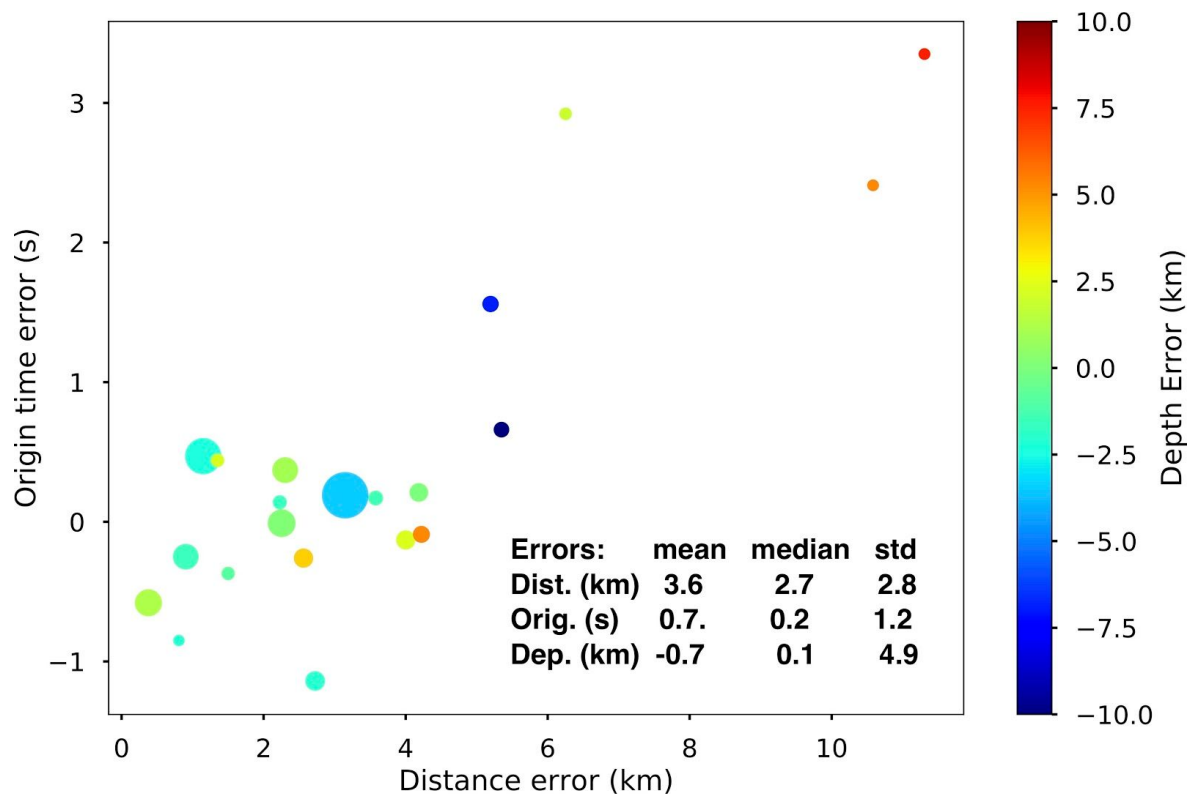

Figure 6. The distance, origin time, and depth error for the 21 events that have good azimuthal coverage and at least 4 phase pickings. The mean, median and standard deviation (std) are also shown for all the errors. The sizes of the circles indicate how many phase pickings were available for each event, which ranged from 4 to 66. See the errors for all the 44 events that have 4 or more phase pickings in Figure S1.

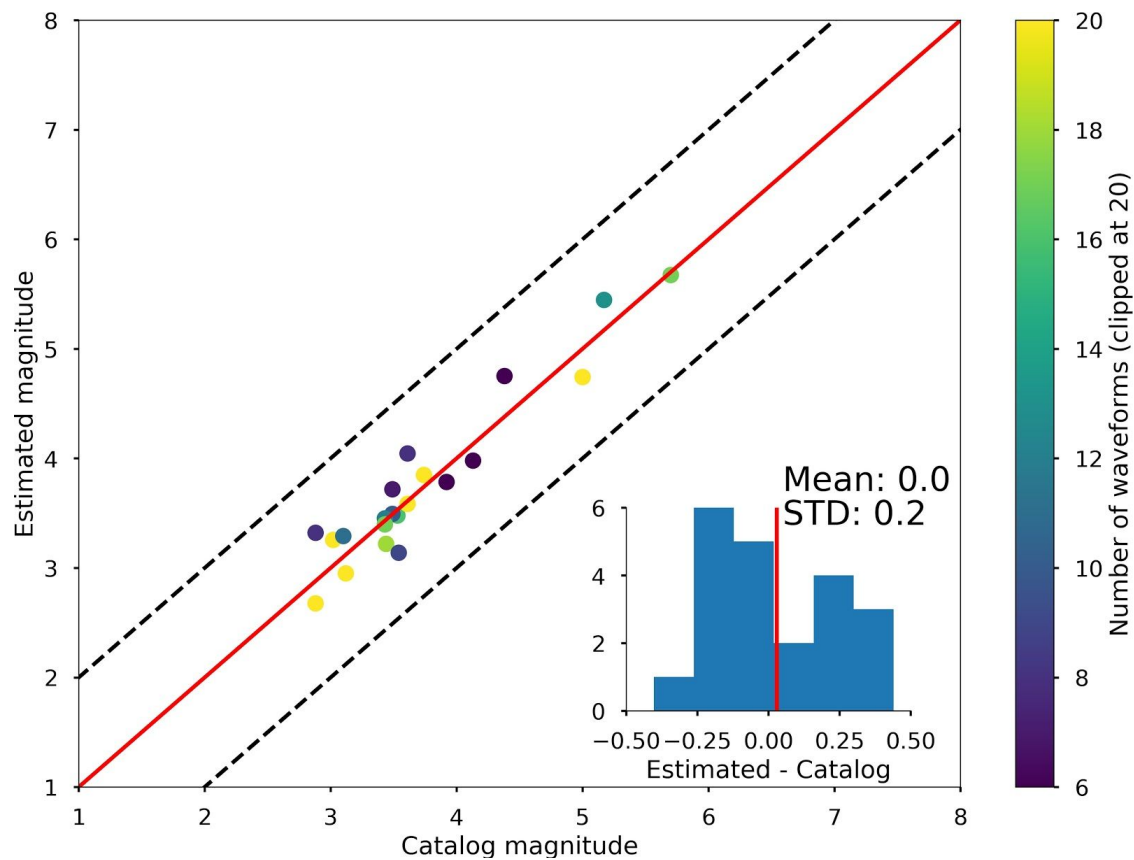

Figure 7. The magnitude estimates from MyShake compared with those from the USGS Comcat. The mean error is 0.0 and standard deviation is 0.2. The color of the circle shows the number of waveforms used. The solid red line is the 1 to 1 line, and two black dotted lines show an error of 1 magnitude unit. The magnitude estimates for all the 44 events that have 4 or more phase pickings are shown in Figure S2.

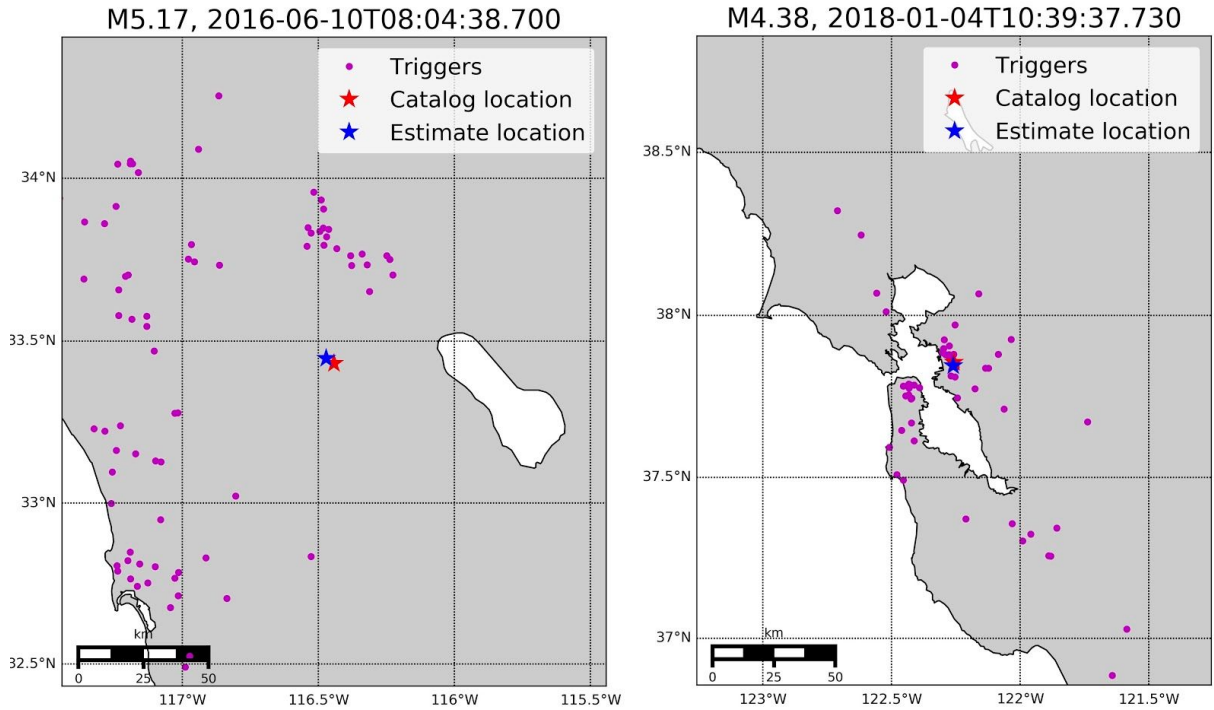

Figure 8. Earthquake source parameter estimation. (a) M5.2 June 10th, 2016 Borrego Springs event, with maximum azimuthal gap of 56 degrees. (b)M4.4 January 4th, 2018 Berkeley event, with maximum azimuthal gap of 17 degrees. The magenta dots are phones that triggered during the earthquake The estimates MyShake epicentral location is shown as a blue star and the USGS catalog location as a red star. The color version of this figure is available only in the electronic edition. Errors of the estimated source parameters with respect to the catalogs are shown in Table 3.

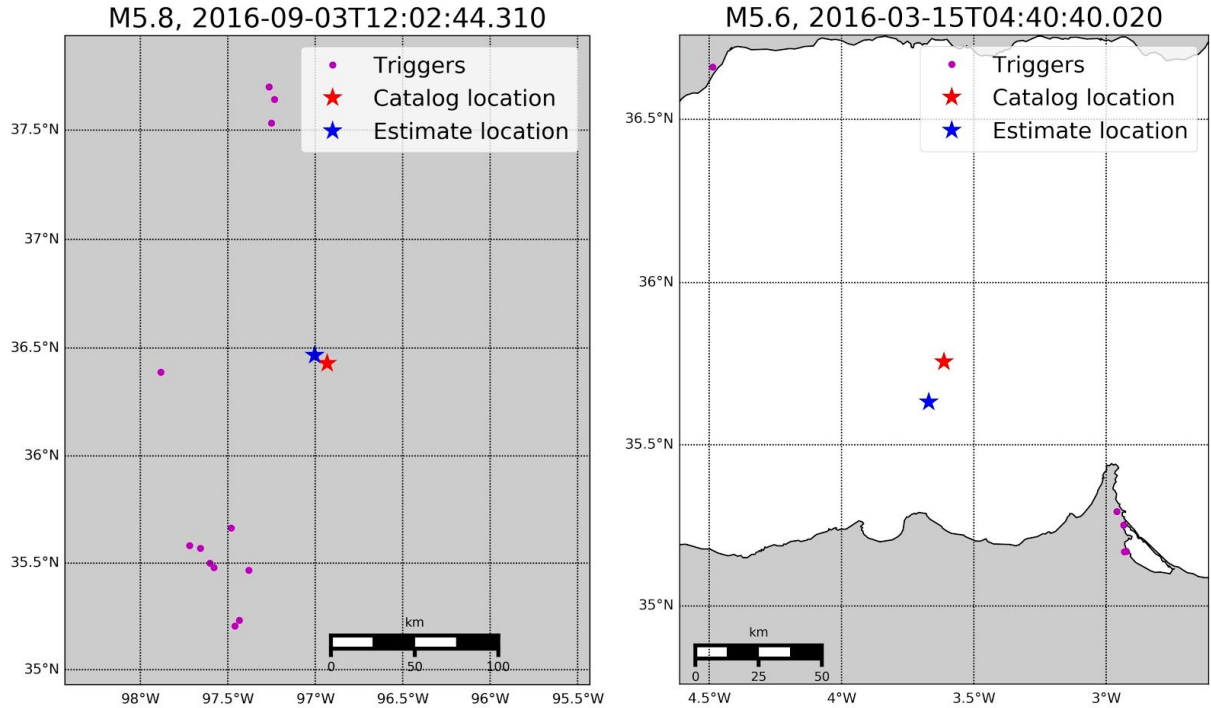

Figure 9. Earthquake source parameter estimation. (a) M5.8 September 3rd, 2016, Pawnee, Oklahoma event, with maximum azimuthal gap of 210 degrees. (b)M5.6 March 15th, 2016 Morocco event, with maximum azimuthal gap of 185 degrees. Note that this event is beneath the Mediterranean Sea offshore of Morocco. The magenta dots are phones that triggered during the earthquake. The estimates MyShake epicentral location is shown as a blue start and the USGS catalog location as a red star. Errors of the estimated source parameters with respect to the catalogs are shown in Table 3.

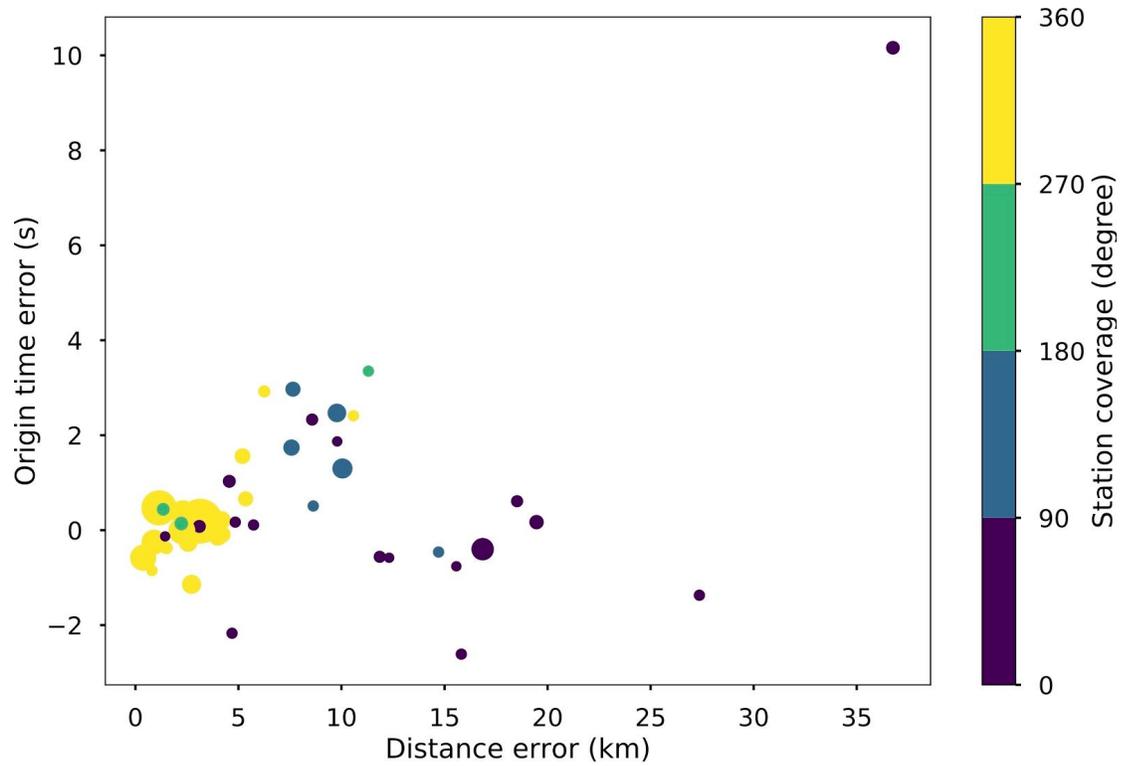

Figure 10. The distance, origin time, and station coverage from Myshake network using all the 44 events that have 4 or more identifiable phases without any coverage filtering. The sizes of the circles indicate how many phase pickings were available for we are using in each event, which ranged from 4 to 66. The colors of the circles showing the station coverage for each event.